\documentclass[aps,pre,twocolumn,superscriptaddress,floatfix]{revtex4-2}

\usepackage{amsmath,amssymb}
\usepackage{bm}
\usepackage{mathptmx}

\usepackage{graphicx}
\usepackage{color}

\usepackage{hyperref}
\usepackage{float}

\usepackage{microtype}
\usepackage{xspace}

\usepackage{makecell}
\usepackage{nicefrac}
\usepackage[normalem]{ulem}

\begin{document}

\title{Dynamic effects of electric fields in a hybrid-coupled thermosensitive neuronal network}

\author{Ediline L.~F. Nguessap}
\affiliation{Department of Physics, FFCLRP, University of S\~ao Paulo, 14040-900 Ribeir\~ao Preto, SP, Brazil}

\author{Antonio C.~Roque}
\affiliation{Department of Physics, FFCLRP, University of S\~ao Paulo, 14040-900 Ribeir\~ao Preto, SP, Brazil}

\author{Fernando F.~Ferreira}
\email{ferfff@usp.br}
\affiliation{Department of Physics, FFCLRP, University of S\~ao Paulo, 14040-900 Ribeir\~ao Preto, SP, Brazil}

\begin{abstract}
The dynamics of thermosensitive FitzHugh--Nagumo neuronal networks under hybrid synaptic coupling and external electric fields are investigated in a ring topology. Numerical simulations show collective behavior ranging from incoherent activity to coherent traveling waves and chimera and multichimera states, arising from the interplay between thermosensitivity, intrinsic electric fields, and the balance between electrical and chemical synapses. Chemical nonlocal coupling is essential for the stabilization of traveling waves and traveling chimera patterns, whereas purely electrical local coupling of chaotic neurons yields only incoherent states. The frequency and spatial extent of the applied field act as key control parameters: low frequencies and localized stimulation can induce or suppress chimera-like domains, while high frequencies have negligible impact on network dynamics. Intrinsic electric fields, controlled by the cell radius $r$, modulate single-neuron excitability and reshape the network response to external fields, enabling transitions between incoherence, chimera-like states, and global synchronization. These results suggest that weak, spatially targeted electric fields can serve as effective control knobs for complex patterns of activity in thermosensitive hybrid-coupled networks, with potential implications for selective neuromodulation and bio-inspired computing.
\end{abstract}

\maketitle

\section{Introduction}

The comprehension of brain function and collective neural dynamics remains a central challenge in contemporary neuroscience. The nervous system exhibits intricate spatiotemporal activity emerging from the interaction between neurons and glial cells, particularly astrocytes, which play a key role in signal exchange and information processing \cite{haydon2006astrocyte,tang2017astrocyte}. Neurons are the basic computational units of the brain, and their electrical activity arises from transmembrane ionic currents carried by calcium, potassium, and sodium ions \cite{hahn2001bistability,gu2014potassium}. To describe these processes, a wide range of theoretical neuron models has been proposed over the past decades \cite{oja1982simplified,rall1962electrophysiology,nagumo1972response,achard2006complex,tsodyks1996population}. In these frameworks, neuronal electrophysiology is usually characterized by the time series of the membrane potential, whose firing patterns are shaped by external stimuli and synaptic inputs \cite{nowotny2007dynamical,colwell2011linking,zandi2020different}. At the network level, various architectures such as rings, chains, small-world, and multilayer structures have been employed to investigate how connectivity and coupling mechanisms (chemical or electrical) influence the emergence of partial or complete synchronization \cite{rakshit2019transitions,xiao2016spatiotemporal,qin2018field}.

A particularly striking phenomenon in complex networks is the appearance of chimera states, in which coherent (synchronized) and incoherent (desynchronized) domains coexist in space. These states were first reported in nonlocally coupled phase oscillators by Kuramoto and co-workers \cite{kuramoto2002coexistence} and later named ``chimera'' in \cite{abrams2004chimera}. Since then, chimera and chimera-like patterns have been identified in a variety of physical, chemical, and biological systems, including neuronal networks \cite{majhi2016chimera,majhi2017chimera,kundu2018chimera,chouzouris2018chimera,andreev2019chimera,majhi2019chimera}. The coexistence of synchronized and desynchronized activity has been suggested to underlie phenomena such as unihemispheric sleep and certain epileptic patterns, which makes chimera states particularly relevant for understanding pathological and functional brain rhythms.

Neuronal activity, governed by transmembrane ion dynamics, is also known to be highly sensitive to weak electric fields \cite{bikson2004effects,radman2007spike}. A landmark contribution by Ma \textit{et al.} \cite{ma2019model} provided a modeling framework to incorporate electric-field effects into neuronal dynamics, revealing a rich repertoire of field-induced behaviors, including bursting synchronization and chimera states \cite{simo2021chimera,wang2016spatiotemporal}. Subsequent works extended this paradigm by introducing additional biophysical mechanisms. For example, Lv and Ma investigated magnetic flux effects via memristive coupling \cite{lv2016multiple,bao2019hidden}, while Liu \textit{et al.} studied photosensitive neural responses \cite{liu2020new}. More recently, thermosensitive FitzHugh--Nagumo (FHN) neurons were shown to undergo temperature-dependent transitions between bursting and chaotic regimes, though these studies focused mainly on single-cell dynamics \cite{xu2020dynamics}. Building on this foundation, recent work demonstrated that thermosensitive FHN neurons subjected to external electric fields can exhibit chaotic bursting and periodic spiking, with the cell size parameter $r$ and the temperature coefficient $b$ acting as key modulators of excitability \cite{nguessap2026modulation}.

Thermosensitive neurons, which respond to temperature variations, have been effectively modeled by coupling a standard FHN oscillator to a thermistor-like element, thereby capturing temperature-dependent excitability and complex firing patterns \cite{xu2020dynamics}. This modeling framework has motivated a series of studies on thermosensitive neural systems in both single-neuron and network settings. Chimera states and synchronization phenomena have been reported in thermosensitive FHN networks \cite{hussain2021chimera,zhu2021effects,guo2022desynchronization}, while other works have examined energy balance \cite{zhou2022energy} and pattern formation in coupled thermosensitive circuits \cite{xu2022pattern}. The versatility of the thermosensitive FHN model has also allowed the incorporation of additional physical influences, such as combined photo-thermal stimulation \cite{tagne2022bifurcations,fossi2022phase}, light-driven modulation \cite{yang2024wave,song2024light}, and electromagnetic induction \cite{ji2024fast}. Collectively, these studies highlight the rich dynamical repertoire of thermosensitive neurons and their networks, as well as the importance of external physical fields as control parameters.

Despite these advances, several fundamental issues remain open. First, the combined impact of thermosensitivity, hybrid synaptic coupling (electrical plus chemical), and externally applied electric fields has not yet been systematically explored in networked systems, even though these elements naturally coexist in biological neural tissue. This gap is critical because temperature-dependent ion channel kinetics can substantially modify how networks respond to field-induced synchronization and desynchronization. Second, while intrinsic electric fields, modeled through a cell-size-dependent variable $r$ are known to shape single-neuron chaotic dynamics, their interaction with external fields at the network level remains largely unknown, even though endogenous field effects arising from cell morphology and externally applied fields are both present in vivo. Third, existing strategies for controlling chimera states typically lack spatial specificity, which limits their applicability to neuromodulation scenarios where localized desynchronization is clinically desirable, such as focal epilepsy therapy.

In this work, these gaps are addressed by investigating a thermosensitive FHN neuronal network with hybrid synaptic coupling under the influence of an external electric field. The model consists of a ring of thermosensitive FHN neurons in which each node includes an intrinsic electric field variable coupled to the membrane dynamics via the cell size parameter $r$. Neurons interact through a hybrid scheme combining local electrical (diffusive) coupling and nonlocal chemical (synaptic) coupling, capturing the complementary roles of gap junctions and synaptic transmission in real neural tissue. An externally applied, periodic electric field is introduced as a spatially localized and tunable modulation, enabling the systematic study of frequency- and space-dependent control of the network dynamics. Within this framework, extensive numerical simulations reveal three main classes of emergent behavior: incoherent activity, coherent traveling waves, and chimera or multichimera states.

Our results show that chemical coupling is essential for the emergence and stabilization of traveling waves and traveling chimera patterns, whereas purely electrical nearest-neighbor coupling of chaotic thermosensitive neurons leads to incoherent network states over a broad range of coupling strengths. The intrinsic electric field, modulated by the cell radius $r$, plays a dual role: it controls the excitability and chaoticity of individual neurons and, at the same time, reshapes the collective response of the network to external electric fields. Furthermore, the frequency and spatial extent of the external field act as key control parameters that organize the collective dynamics. Low-frequency, localized stimulation can induce robust chimera-like states by creating adjacent coherent and incoherent domains, or it can suppress chaotic activity and promote synchronization, while high-frequency forcing tends to have negligible impact on the global dynamics. By systematically mapping these regimes in parameter space, the study establishes how thermosensitivity, intrinsic and extrinsic electric fields, and hybrid synaptic architecture cooperate to generate and control complex spatiotemporal patterns.

Overall, the proposed framework demonstrates that weak, spatially targeted electric fields can function as effective levers for pattern selection in thermosensitive hybrid-coupled networks. This provides mechanistic insight into how external stimulation might be used to engineer or suppress chimera-like structures and traveling waves in neural tissue, with potential implications for the design of more selective neuromodulation strategies and bio-inspired computing architectures.


\section{Model}
\label{sec:model}

We consider a thermosensitive FitzHugh--Nagumo (FHN) neuron extended with an intrinsic electric-field variable and driven by an external periodic current\cite{nguessap2026modulation}. The single-neuron dynamics are
\begin{align}
\frac{dx}{dt} &= x - \frac{1}{3}x^{3} - y + I + A \cos(\omega t),
\label{eq:single_x}\\
\frac{dy}{dt} &= c\bigl[x + a - b \exp(1/T) - y\bigr] - r E,
\label{eq:single_y}\\
\frac{dE}{dt} &= k y,
\label{eq:single_E}
\end{align}
where $x$ is the membrane potential, $y$ is a recovery (ionic current) variable, and $E$ is the intrinsic electric field. The parameters $I$ and $A$ are the constant and time-periodic components of the external current with angular frequency $\omega$  respectively. The parameters $a$ and $b$ control excitability and thermosensitivity, $T$ is a reference temperature, $c$ sets the time scale of $y$, $k$ scales the intrinsic field, and $r$ r is a dimensionless parameter representing the effective strength of the intrinsic electric-field coupling. The variable $E$ provides a phenomenological description of ephaptic field effects generated by transmembrane ionic currents in the near extracellular space.
\begin{figure}[htp]
\centering
    \centering
    \begin{tabular}{cc}
       (a) & (b)\\
        \includegraphics[width=0.235\textwidth]{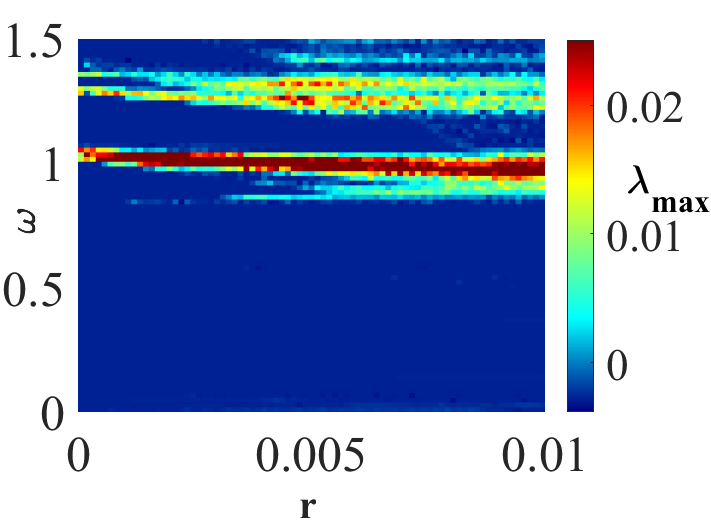} &
      \includegraphics[width=0.235\textwidth]{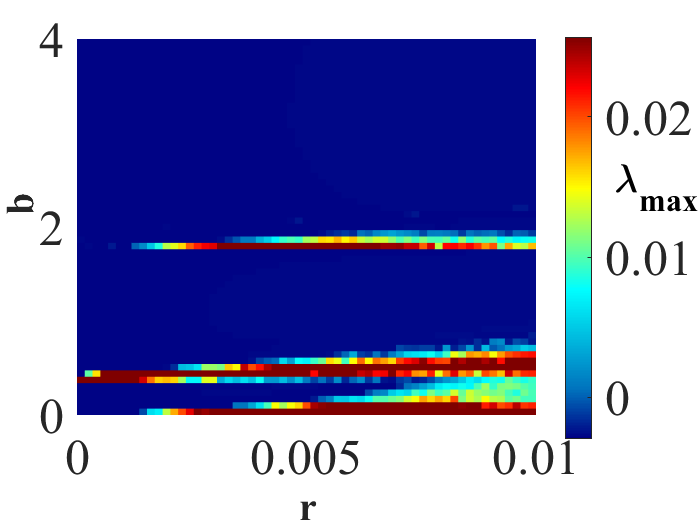}\\

    \end{tabular}
  \caption{ Parameter-space analysis of the single-neuron dynamics characterized by the largest Lyapunov exponent $\lambda_{\max}$. (a) $\lambda_{\max}$ as a function of driving frequency $\omega$ and cell radius $r$ at fixed $b=0.4$. (b) $\lambda_{\max}$ as a function of $b$ and $r$ at fixed $\omega=1.004$. Blue regions ($\lambda_{\max}\approx0$) correspond to periodic activity, whereas regions with $\lambda_{\max}>0$ correspond to chaotic firing. Other parameters: $A=0.9$, $a=0.7$, $c=0.1$, $\xi=0.175$, $I=0.5$, $T=5$, and $k=0.001$.
}
  \label{fig:1}
  \end{figure}
  
  \begin{figure}[htp]
\centering
    \centering
    \begin{tabular}{cc}
       (a) & (b)\\
       \includegraphics[width=0.235\textwidth]{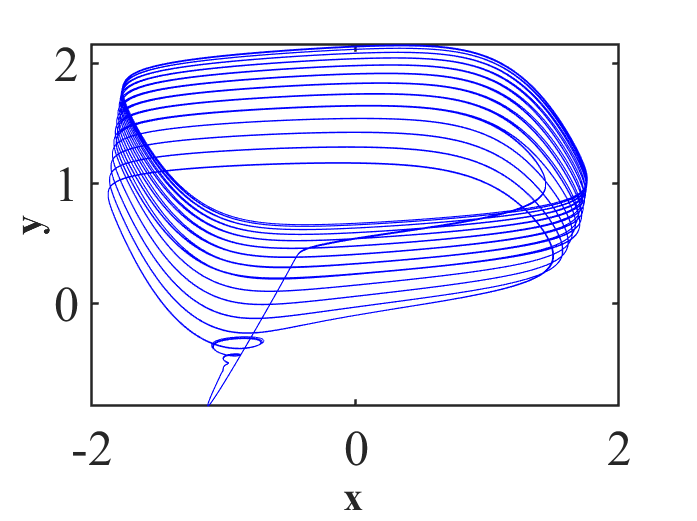} &  
       \includegraphics[width=0.235\textwidth]{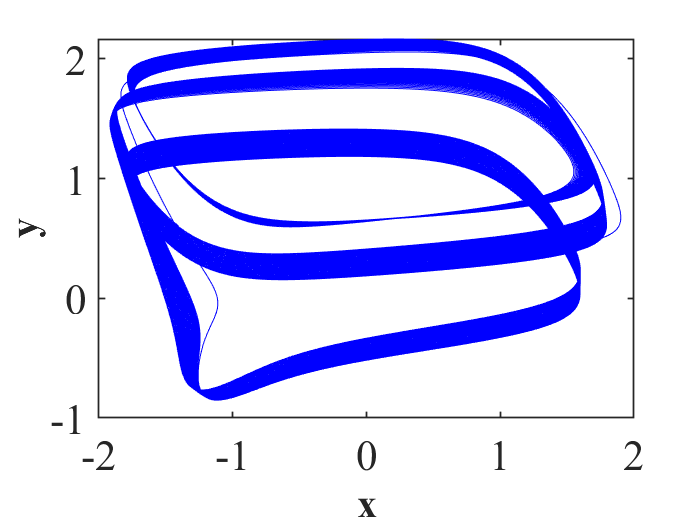}\\
       
       (c) & (d)\\
       \includegraphics[width=0.235\textwidth]{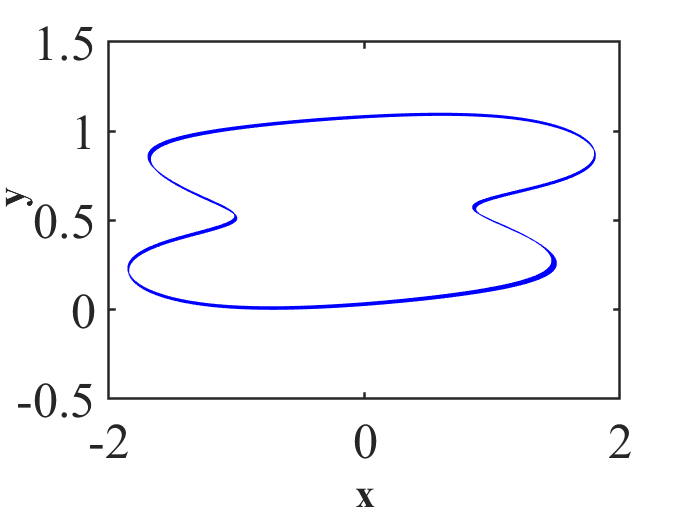}&
       \includegraphics[width=0.235\textwidth]{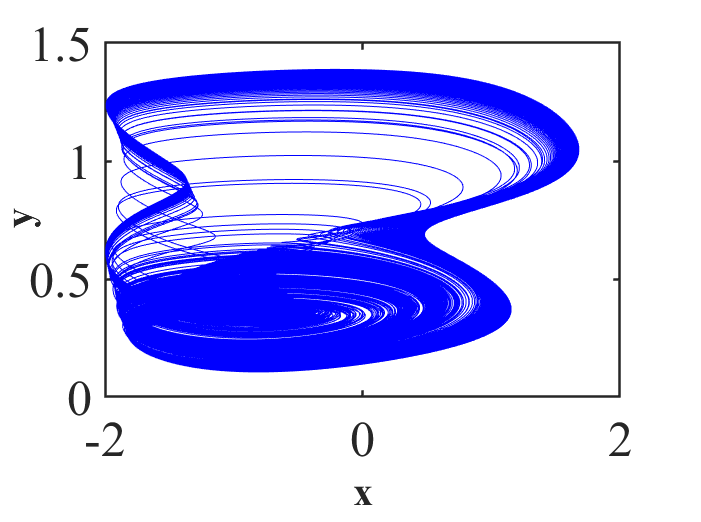}
        
    \end{tabular}
    \label{tab:placeholder}

  \caption{Phase diagrams of the single neuron under periodic forcing. (a) Periodic bursting for $b=0.4$, $\omega=0.005$, $r=10^{-4}$. (b) Periodic bursting for $b=0.4$, $\omega=0.05$, $r=0.01$. (c) Periodic spiking for $b=0.8$, $\omega=1.004$, $r=10^{-4}$. (d) Chaotic spiking for $b=0.4$, $\omega=1.004$, $r=0.007$.}
  \label{fig:2}
\end{figure}

To study collective behavior, we arrange $N$ identical neurons in a ring and include hybrid synaptic coupling and an externally applied electric field, following Ref.~\cite{simo2021chimera}. The dynamics of neuron $i$ are
\begin{equation}
\begin{cases}
\displaystyle
\frac{dx_i}{dt} =
x_i(1-\xi) - \frac{1}{3}x_i^3 - y_i + I + A\cos(\omega t)
+ J_i + C_i,\\[6pt]
\displaystyle
\frac{dy_i}{dt} =
c\bigl[x_i + a - b\exp(1/T) - y_i\bigr] + r E_i,\\[6pt]
\displaystyle
\frac{dE_i}{dt} =
k y_i + E_{\text{ext}}(t),
\end{cases}
\label{eq:network}
\end{equation}
where $\xi$ is a parameter of the modified FHN dynamics and $E_{\text{ext}}(t)$ is an external electric field.

The external field is a sinusoidal signal
\begin{equation}
E_{\text{ext}}(t) = E_m \sin(2\pi f t),
\end{equation}
with amplitude $E_m$ and frequency $f$.

Electrical coupling between nearest neighbors is
\begin{equation}
J_i = d\bigl(x_{i+1} + x_{i-1} - 2x_i\bigr),
\end{equation}
where $d$ is the electrical coupling strength and periodic boundary conditions are assumed.

Nonlocal chemical coupling is given by
\begin{equation}
C_i = \frac{\epsilon}{2p-2}\left(x_i-xs\right)
      \left[
        \sum_{j=i-p}^{i+p} \Gamma(x_j)
        - \sum_{j=i-1}^{i+1} \Gamma(x_j)
      \right],
\end{equation}
with
\begin{equation}
\Gamma(x_j) = \frac{1}{1+e^{-\Lambda(x_j-\theta_s)}},
\end{equation}
where $\epsilon$ is the chemical coupling strength, $x_s=2$ is the reversal potential, $\theta_s=-0.25$ is the synaptic threshold, $\Lambda=10$ sets the steepness, and $p$ is the nonlocal coupling range. This hybrid scheme reproduces the coexistence of local electrical and nonlocal chemical communication and matches previous work on chimera states in neuronal networks \cite{simo2021chimera,hussain2021chimera}. Unless stated otherwise, we use $a=0.7$, $c=0.1$, $\xi=0.175$, $I=0.5$, $T=5$, $k=0.001$, and initial conditions $(x_0,y_0,E_0)=(0.1,0.3,0.003)$.

\section{Single-neuron dynamics}
\label{sec:single}

The single-neuron model defined by Eqs.~\eqref{eq:single_x}–\eqref{eq:single_E} exhibits a rich variety of dynamical regimes, including periodic bursting, periodic spiking, and chaotic spiking, depending on the driving frequency $\omega$, the thermosensitive parameter $b$, and the intrinsic electric-field coupling parameter $r$. To characterize these regimes, we computed the largest Lyapunov exponent $\lambda_{\max}$ using the standard algorithm based on the evolution and periodic renormalization of a tangent perturbation vector, following the approaches introduced by Benettin \textit{et al.}~\cite{benettin1980lyapunov} and Wolf \textit{et al.}~\cite{wolf1985determining}.

Figure~\ref{fig:1} presents the largest Lyapunov exponent in two parameter planes. In Fig.~\ref{fig:1}(a), $b=0.4$ is fixed while the driving frequency $\omega$ and the cell-radius parameter $r$ are varied. In Fig.~\ref{fig:1}(b), the driving frequency is fixed at $\omega=1.004$ while $b$ and $r$ are varied. The blue regions, characterized by $\lambda_{\max}\approx0$, correspond to periodic activity, whereas the remaining regions with $\lambda_{\max}>0$ correspond to chaotic dynamics. The results indicate that both the thermosensitive parameter $b$ and the intrinsic electric-field coupling parameter $r$ strongly affect the transition between periodic and chaotic firing.

The results reveal that the interplay between thermosensitivity, external forcing, and intrinsic electric-field feedback strongly affects neuronal excitability. Increasing the intrinsic field coupling parameter $r$ generally enlarges the regions of irregular activity, while changes in the thermosensitive parameter $b$ shift the boundaries between periodic and chaotic regimes. Likewise, the driving frequency $\omega$ can induce transitions between regular and irregular firing patterns.

Representative phase portraits corresponding to different regions of Fig.~\ref{fig:1} are shown in Fig.~\ref{fig:2}. For $b=0.4$, $\omega=0.005$, and $r=10^{-4}$, the neuron exhibits periodic bursting dynamics (Fig.~\ref{fig:2}(a)). For $b=0.4$, $\omega=0.05$, and $r=0.01$, the neuron remains in a periodic bursting regime, although with a different oscillatory structure (Fig.~\ref{fig:2}(b)). Increasing the thermosensitive parameter to $b=0.8$ while keeping $\omega=1.004$ and $r=10^{-4}$ produces periodic spiking activity (Fig.~\ref{fig:2}(c)). In contrast, for $b=0.4$, $\omega=1.004$, and $r=0.007$, the neuron displays chaotic spiking characterized by an irregular trajectory in phase space (Fig.~\ref{fig:2}(d)).

These results demonstrate that the intrinsic electric field and thermosensitive mechanisms jointly regulate the transition between bursting, periodic spiking, and chaotic firing. The identified dynamical regimes serve as the basis for the network investigations presented in the following sections.


\section{Network dynamics and metrics}
\label{sec:network}

We analyze the dynamics of the thermosensitive FitzHugh--Nagumo network under different coupling configurations and external electric fields.  All simulations are performed using a fourth-order Runge--Kutta scheme with a fixed time step $\Delta t = 0.01$ and a total of $2 \times 10^5$ iterations. To eliminate transient effects, the initial $60\%$ of the simulation is discarded. The remaining time series is re-indexed such that $t = 1$ corresponds to the beginning of the post-transient regime. The variable $t$ denotes the discrete iteration index; since $\Delta t = 0.01$, the iteration index is proportional to physical time via $t_{\text{phys}} = t \Delta t$. Unless otherwise specified, all snapshots shown in this work are taken at time index $t = 5000$ within the post-transient regime.

To characterize local coherence we use a Kuramoto-type order parameter \cite{omelchenko2013nonlocal}. The geometric phase of neuron $k$ is
\begin{equation}
\Phi_k = \arctan\left(\frac{y_k}{x_k}\right),
\end{equation}
and the local order parameter at node $i$ is
\begin{equation}
L_i = \left|\frac{1}{2\eta}\sum_{|i-k|\leq\eta} e^{i\Phi_k}\right|,
\end{equation}
where $\eta$ is the number of neighbors in the local average; $L_i\approx 1$ indicates local coherence and $L_i\approx 0$ indicates local incoherence.
In addition to the local order parameter, we also compute the global Kuramoto order parameter to quantify overall phase coherence across the network. It is defined as
\begin{equation}
R = \left| \frac{1}{N} \sum_{k=1}^{N} e^{i \Phi_k} \right|,
\end{equation}
where $\Phi_k$ is the geometric phase defined in Eq.~(9). The value $R \approx 1$ indicates global synchronization, whereas $R \approx 0$ corresponds to incoherent dynamics. Intermediate values of $R$ are typically associated with partially synchronized states such as chimera patterns.
Global coherence and chimera structure are quantified by the strength of incoherence (SI) and the discontinuity measure (DM) \cite{gopal2014observation}. Defining nearest-neighbor differences $z_i=x_{i+1}-x_i$ and partitioning the ring into $M$ bins of size $n=N/M$, the local standard deviation in bin $m$ is
\begin{equation}
\sigma(m) = \left\langle 
\sqrt{\frac{1}{n}\sum_{j=n(m-1)+1}^{nm}\bigl(z_j - \langle z\rangle_m\bigr)^2} 
\right\rangle_t,
\end{equation}
and a bin is classified as coherent if $\sigma(m)<\delta$, with $\delta$ a small threshold. The strength of incoherence is
\begin{equation}
\mathrm{SI} = 1 - \frac{1}{M}\sum_{m=1}^{M} S_m,
\qquad
S_m = \Theta\bigl(\delta - \sigma(m)\bigr),
\end{equation}
so that SI$=0$ for fully coherent states, SI$=1$ for fully incoherent states, and $0<\mathrm{SI}<1$ for chimera-like states. The discontinuity measure
\begin{equation}
\mathrm{DM} = \frac{1}{2}\sum_{m=1}^{M}\bigl|S_m - S_{m+1}\bigr|,
\qquad S_{M+1}\equiv S_1,
\end{equation}
counts transitions between coherent and incoherent bins; DM$=1$ is typical of single chimera states, whereas DM$>1$ indicates multichimera patterns.
 \begin{figure*}[htp]
\centering
   
    \begin{tabular}{ccc}
        (a) & (b) & (c)\\
        \includegraphics[width=0.235\textwidth]{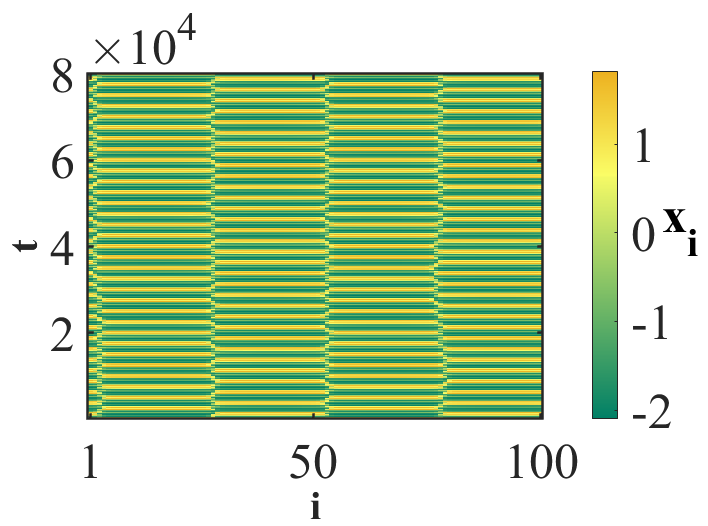} &  
       \includegraphics[width=0.235\textwidth]{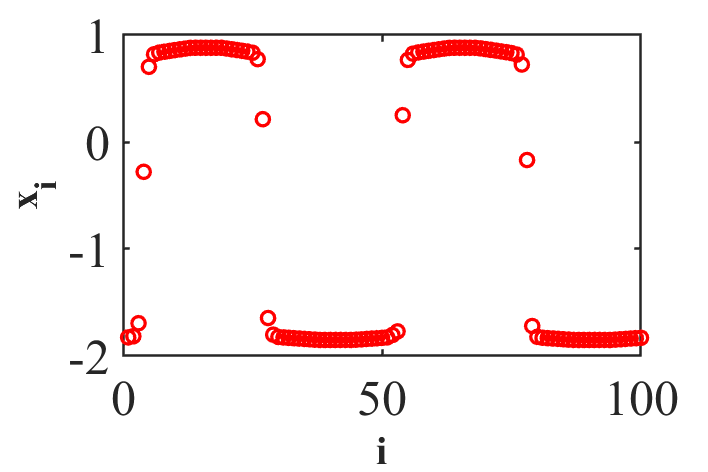}&      
       \includegraphics[width=0.235\textwidth]{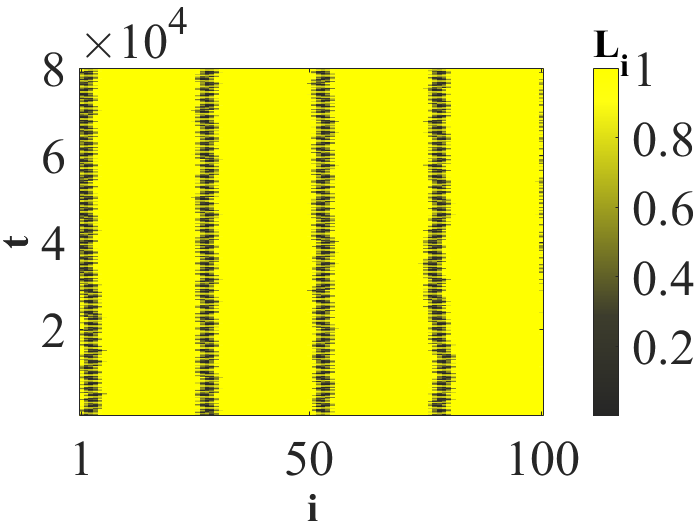}\\
       (c) & (d) & (e)\\
       \includegraphics[width=0.235\textwidth]{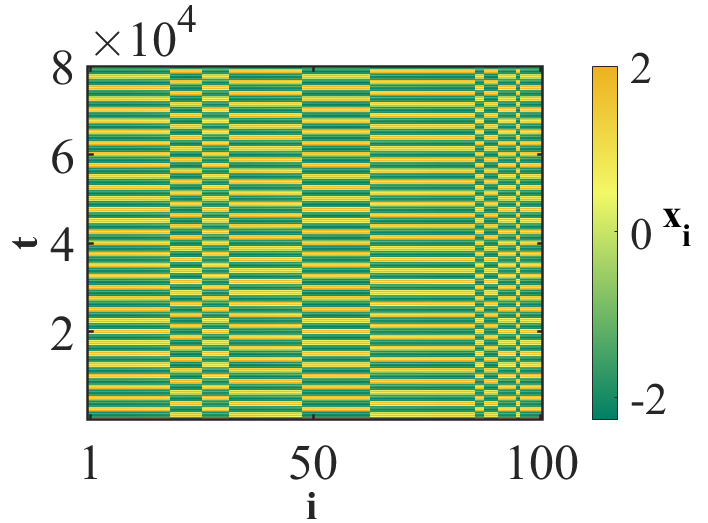} &
       \includegraphics[width=0.235\textwidth]{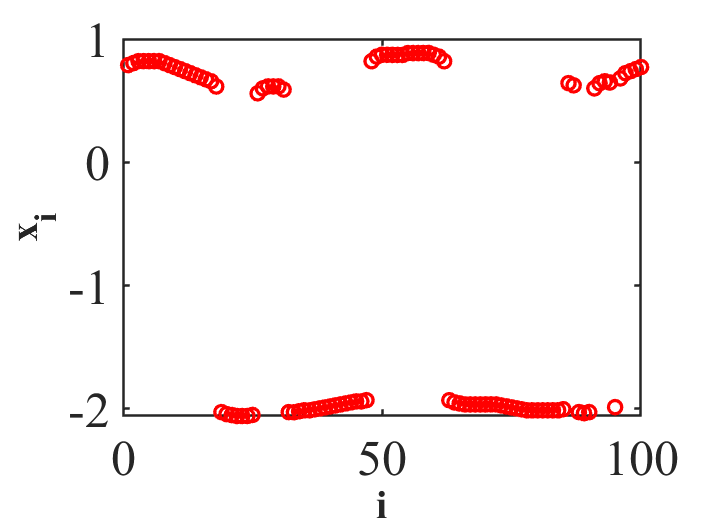}&
       \includegraphics[width=0.235\textwidth]{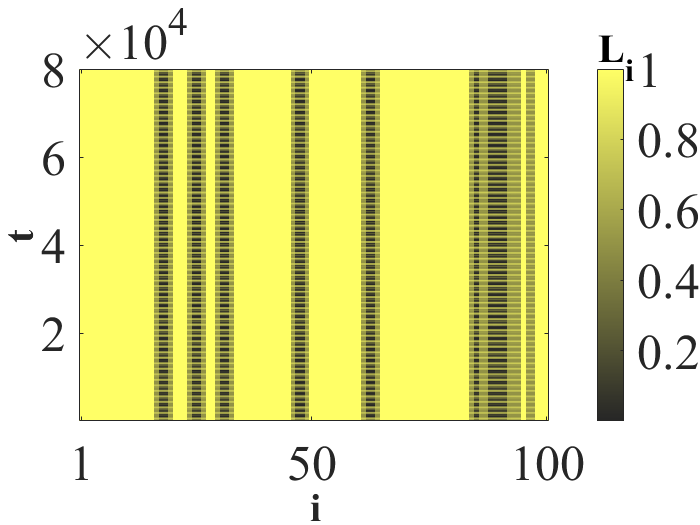}\\
        
    \end{tabular}
    \label{tab:placeholder}
\caption{Chimera-like states for $r=0$ and weak chemical coupling. (a,b) spatiotemporal dynamic and corresponding snapshot  or $\epsilon=0.2$ a few large coherent clusters coexist with incoherent domains.  (d,e) spatiotemporal dynamic and corresponding snapshot  for $\epsilon=0.5$ the pattern fragments into a multicluster chimera with smaller coherent groups. (c,f) Local order parameter $L_i$ with $L_i\approx 1$ in coherent clusters and $L_i\approx 0$ in incoherent regions. Parameters: $a=0.7$, $c=0.1$, $\xi=0.175$, $A=0.9$, $T=5$.Snapshot taken at $t=5000$(post-transient regime)}
\label{fig:3}
 \end{figure*}

Unless stated otherwise, we consider a ring of $N$ identical neurons with
$a=0.7$, $c=0.1$, $\xi=0.175$, $T=5$, $k=0.001$, $b=0.4$, $I=0.5$, $A=0.9$, and $\omega=1.004$. Initial conditions are
\begin{align}
x_i &= 0.1\left(i-\frac{N}{2}\right) + \zeta_{x_i},\\
y_i &= 0.3\left(i-\frac{N}{2}\right) + \zeta_{y_i},\\
E_i &= 0.003\left(i-\frac{N}{2}\right) + \zeta_{E_i}.
\end{align}
where $\zeta_{x_i}$, $\zeta_{y_i}$, and $\zeta_{E_i}$ are small random perturbations and $i=1,\dots,N$. Small uniformly distributed perturbations were added to the initial conditions within the intervals:
$ \zeta_{x_i} \in [0, 10^{-2}], \quad \zeta_{y_i} \in [0, 2 \times 10^{-2}] \quad\text{and}  \quad \zeta_{E_i}=0$ These values are small relative to the state-variable amplitudes and do not alter the qualitative dynamics.

\begin{figure}[htp]
\centering
    \centering
    \begin{tabular}{cc}
       (a) & (b)\\
       \includegraphics[width=0.235\textwidth]{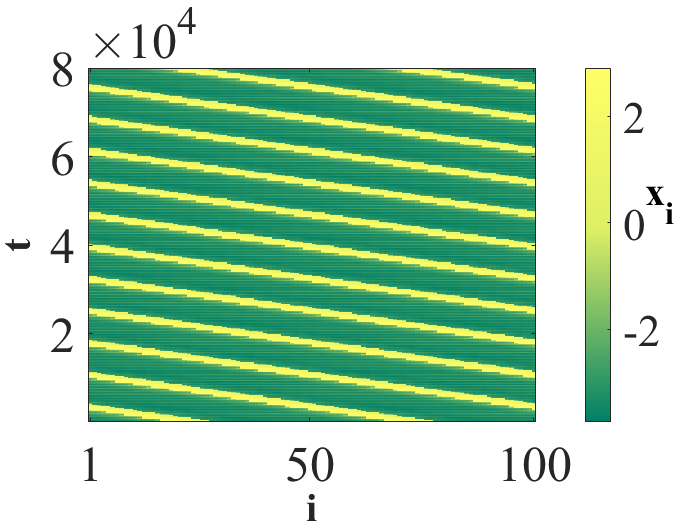} &  
       \includegraphics[width=0.235\textwidth]{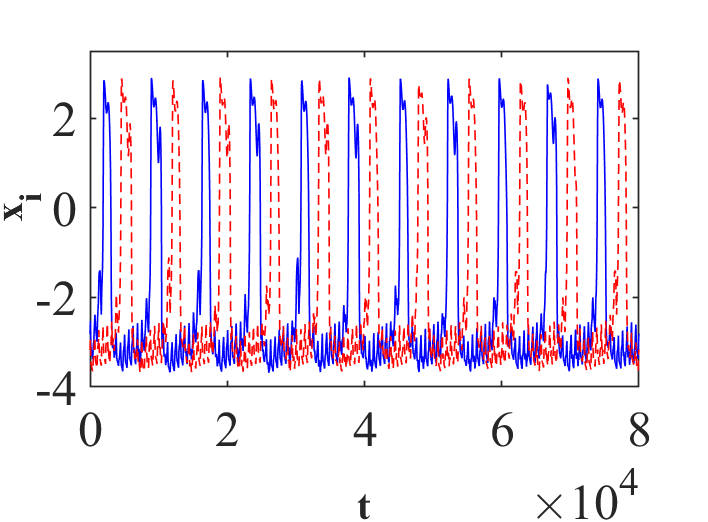}\\
       
       (c) & (d)\\
       \includegraphics[width=0.235\textwidth]{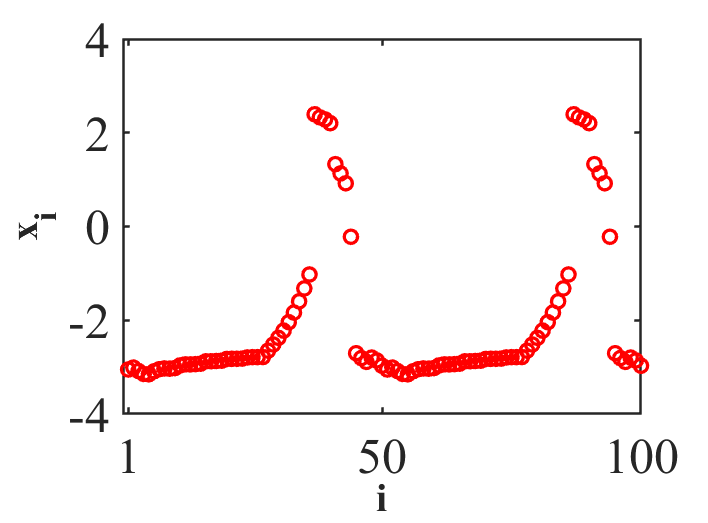}&
       \includegraphics[width=0.235\textwidth]{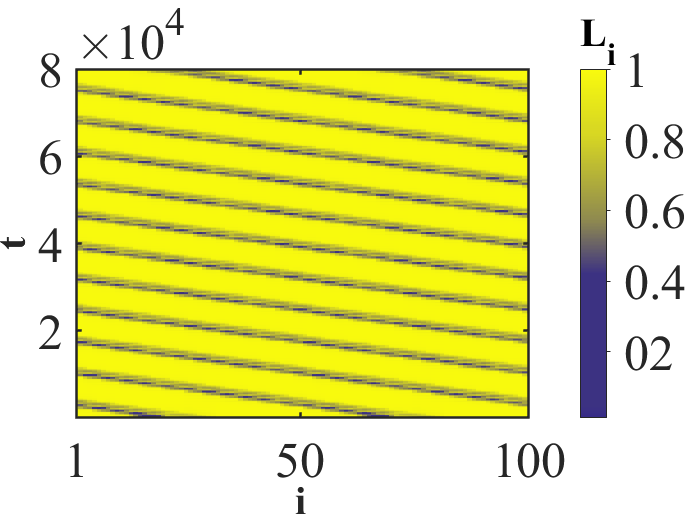}
        
    \end{tabular}
    \label{tab:placeholder}

\caption{Traveling wave for $r=0$, $\epsilon=10$, and $d=0.001$. (a) Spatiotemporal dynamics of $x_i$. (b) Time evolution of $x_i$ for neurons $i=8$ and $i=88$, showing a constant phase lag. (c) Snapshot of $x_i$ at a fixed time, revealing a smooth phase gradient. (d) Local order parameter $L_i\approx 1$ for all nodes, confirming coherent wave propagation. Parameters: $a=0.7$, $c=0.1$, $\xi=0.175$, $A=0.9$, $T=5$, $k=0.001$. Snapshot taken at $t=5000$(post-transient regime)}
\label{fig:4}
 \end{figure}
\subsection{Dynamics without external field}
\label{sec:no_field}

\subsubsection{Case $r=0$}

We first set $r=0$ to suppress the intrinsic electric field effect and focus on the effect of chemical coupling. This case corresponds to the limit where intrinsic electric-field effects are negligible, reducing the model to the classical or standard formulation without field coupling. For weak nonlocal coupling ($\epsilon<1$), the network supports chimera-like patterns. Figure\ref{fig:3} shows two examples for $\epsilon=0.2$ and $\epsilon=0.5$. For $\epsilon=0.2$, the oscillators form a few large coherent clusters separated by incoherent regions [Figs.\ref{fig:3}(a),(b)]; increasing the coupling to $\epsilon=0.5$ fragments the pattern into more, smaller coherent clusters embedded in incoherent domains [Figs.\ref{fig:3}(d),(e)]. The corresponding local order parameter $L_i$ [Figs.\ref{fig:3}(c),(f)] takes values close to one inside coherent clusters and close to zero in the intermediate regions, confirming typical chimera states in this parameter range.

For stronger chemical coupling ($\epsilon>1$), the network exhibits coherent traveling waves. Figure~\ref{fig:4} illustrates a typical case for $\epsilon=10$ and $d=0.001$ at $r=0$. The spatiotemporal map of $x_i$ [Fig.\ref{fig:4}(a)] displays a clear diagonal structure corresponding to a wave propagating around the ring, while the time series of neurons at $i=8$ and $i=88$ [Fig.\ref{fig:4}(b)] show regular oscillations with a fixed phase lag. A snapshot [Fig.\ref{fig:4}(c)] reveals a smooth phase gradient across the network, and $L_i$ remains close to one for all nodes [Fig.~\ref{fig:4}(d)], indicating strong local synchrony consistent with a robust traveling wave.

\begin{figure*}[htp]
\centering   
    \begin{tabular}{ccc}
        (a) & (b) & (c)\\
\includegraphics[width=0.235\textwidth]{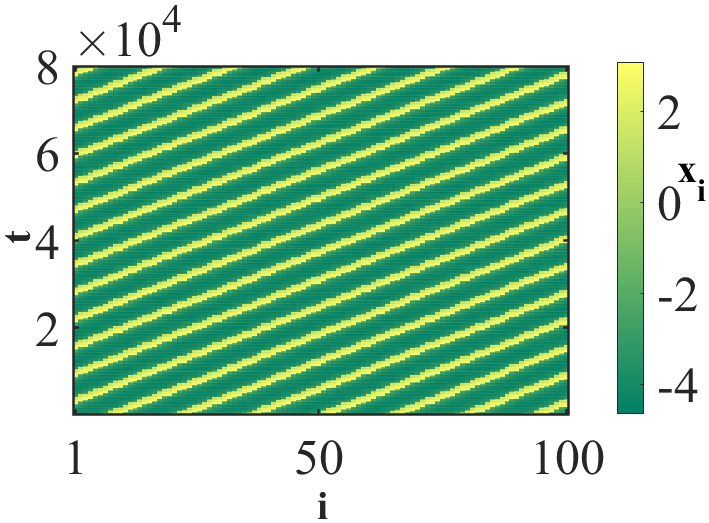} &  
       \includegraphics[width=0.235\textwidth]{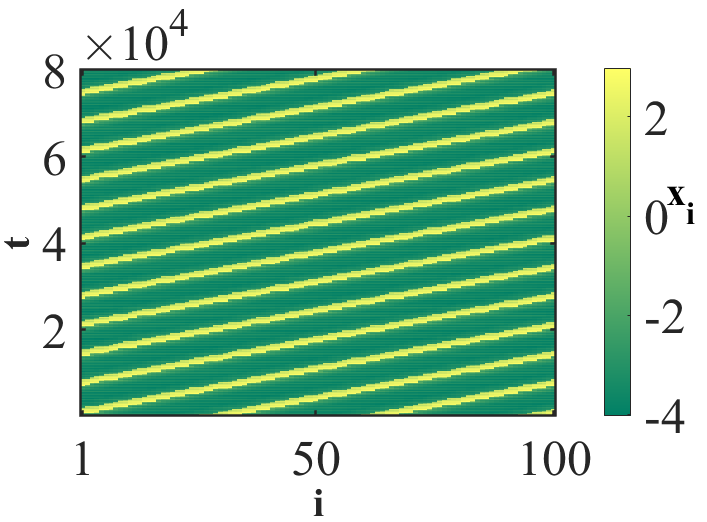} &
      \includegraphics[width=0.235\textwidth]{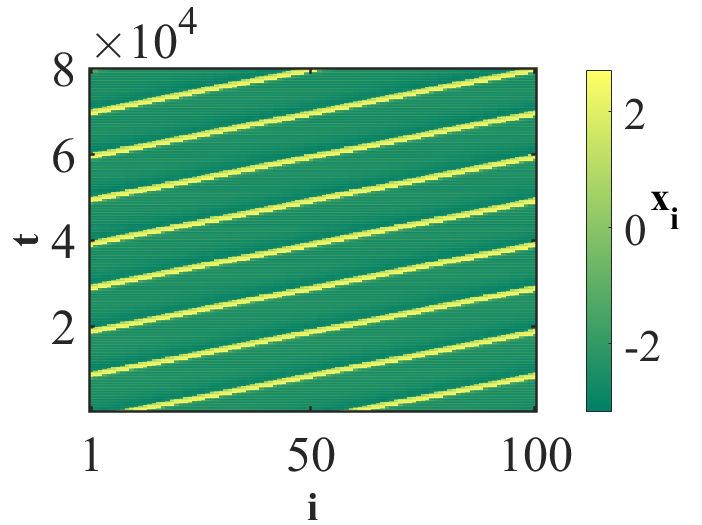}\\
      \end{tabular}
    \label{tab:placeholder}
\caption{Effect of nonlocal coupling range $p$ on traveling waves for $r=0$. Spatiotemporal patterns for (a) $p=10$, (b) $p=25$, and (c) $p=45$ at $\epsilon=9$ and $d=0.001$ show that the traveling wave is preserved, while the number of wave fronts decreases as $p$ increases. Parameters: $a=0.7$, $c=0.1$, $\xi=0.175$, $A=0.9$, $T=5$, $k=0.001$.}
\label{fig:5}
 \end{figure*}

We next examine the influence of the nonlocal coupling range $p$ on these waves. Figure~\ref{fig:5} shows spatiotemporal patterns for $p=10$, $25$, and $45$ at $\epsilon=9$ and $d=0.001$. In all cases the diagonal pattern is preserved, demonstrating that the wave mechanism is robust to changes in $p$; increasing $p$ decreases the number of wave fronts in a fixed time window, indicating a lower initiation frequency and longer inter-burst intervals. This effect is consistent with stronger effective inhibition as each neuron receives input from more neighbors, raising the threshold for wave initiation.

\subsubsection{Case: non-zero $r$}

\begin{figure}[htp]
\centering
    \centering
    \begin{tabular}{cc}
       (a) & (b)\\
       \includegraphics[width=0.235\textwidth]{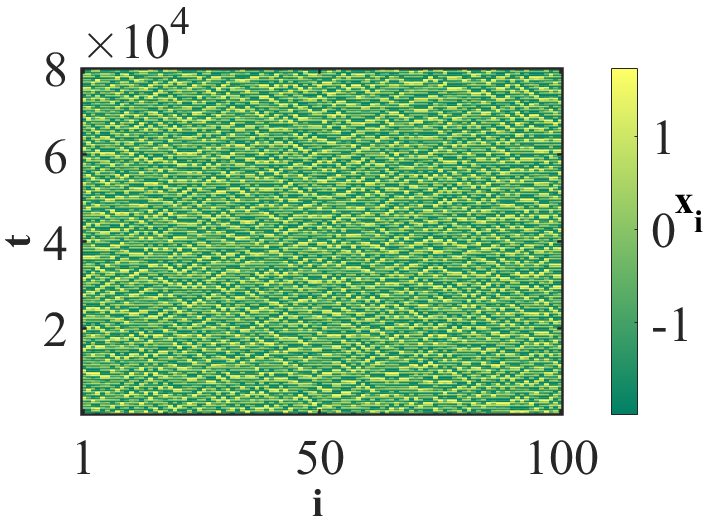} &  
       \includegraphics[width=0.235\textwidth]{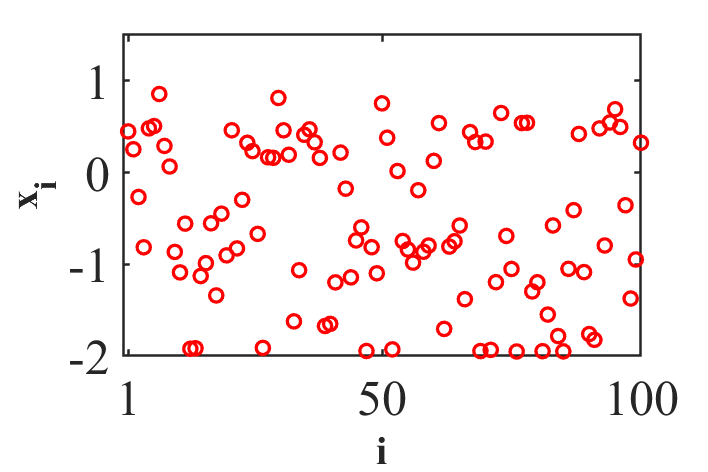}\\
       
       (c) & (d)\\
       \includegraphics[width=0.235\textwidth]{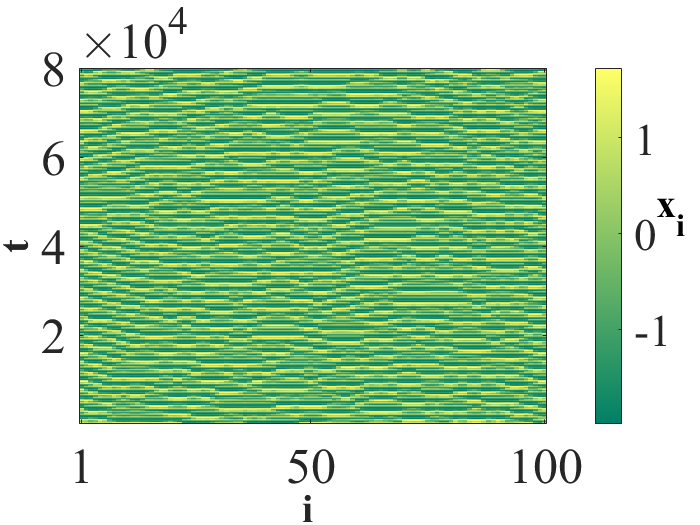}&
       \includegraphics[width=0.235\textwidth]{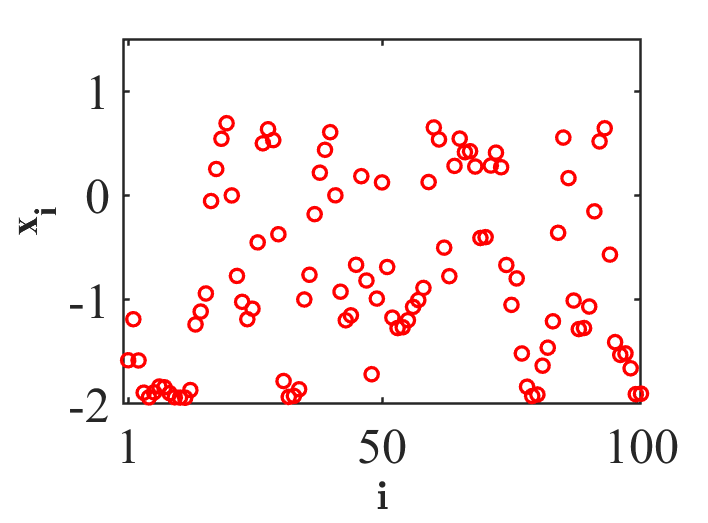}
        
    \end{tabular}
    \label{tab:placeholder}
\caption{Incoherent dynamics for $r=0.007$ and purely electrical coupling ($\epsilon=0$). Spatiotemporal patterns and snapshots for (a,b) $d=0.0001$ and (c,d) $d=0.1$ show spatially disordered, chaotic firing across the network. Parameters: $a=0.7$, $c=0.1$, $\xi=0.175$, $A=0.9$, $T=5$, $k=0.001$, $\epsilon=0$.}
\label{fig:6}
 \end{figure}

We now set $r=0.007$ so that the intrinsic electric field is active and the single neuron operates in a chaotic regime for the chosen parameters (Sec.~\ref{sec:single}). With purely electrical coupling ($\epsilon=0$), the network remains incoherent over a wide range of $d$. Figure~\ref{fig:6} shows spatiotemporal patterns and snapshots of $x_i$ for $d=0.0001$ and $d=0.1$; in both cases neurons fire chaotically and independently, without visible spatial order, indicating complete desynchronization under local electrical coupling of chaotic units.

\begin{figure}[htp]
\centering
    \centering
    \begin{tabular}{cc}
       (a) & (b)\\
        \includegraphics[width=0.235\textwidth]{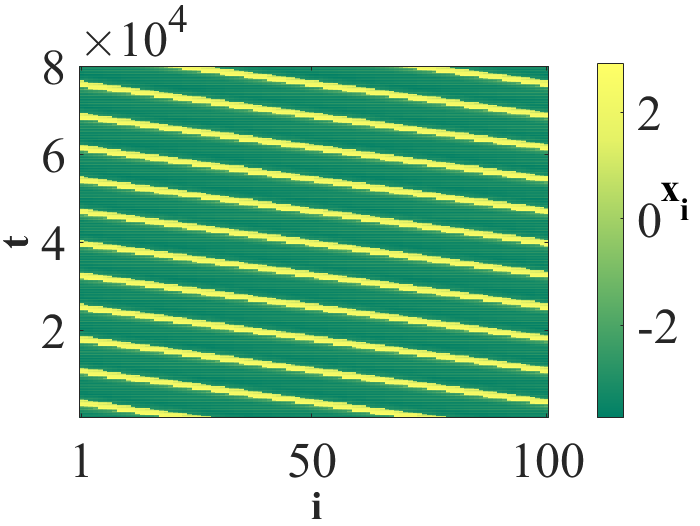} &
      \includegraphics[width=0.235\textwidth]{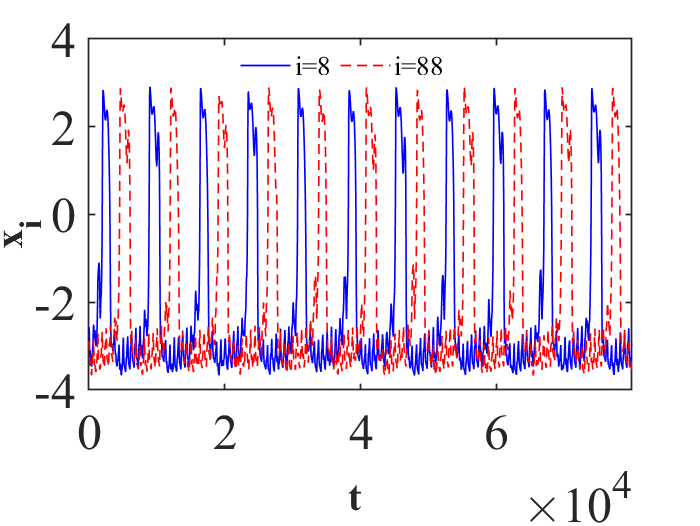}\\

    \end{tabular}
    \label{tab:placeholder}

\caption{Traveling wave pattern for $r=0.007$ with chemical coupling. (a) Spatiotemporal evolution of $x_i$ for $\epsilon=10$; the color bar indicates the magnitude of $x_i$. (b) Time series of $x_i$ for neurons $i=8$ and $i=88$, showing phase-locked oscillations. Parameters: $a=0.7$, $c=0.1$, $\xi=0.175$, $A=0.9$, $T=5$, $k=0.001$, $\omega=1.004$, $p=40$, $d=0.0001$.}
\label{fig:8}
 \end{figure}

When nonlocal chemical coupling is restored with sufficiently large $\epsilon$, traveling waves reappear despite the chaotic single-neuron dynamics. Figure~\ref{fig:8} shows a representative case for $r=0.007$, $\epsilon=10$, $d=0.0001$, and $p=40$; the spatiotemporal pattern [Fig.~\ref{fig:8}(a)] displays a coherent traveling wave, and the time series of neurons $i=8$ and $i=88$ [Fig.~\ref{fig:8}(b)] show regular oscillations with a fixed phase shift. This confirms that strong nonlocal chemical coupling can impose and stabilize wave propagation in a network of chaotic thermosensitive neurons.

\subsection{Dynamics with external electric field}
\label{sec:field}

We now study the effect of an external periodic electric field on the chaotic network with $r=0.007$. Following Ref.~\cite{simo2021chimera}, a weak sinusoidal field of amplitude $E_m=1.5$ is applied to a subset of $N_E$ neurons.

\subsubsection{Only electrical coupling}

\begin{figure}[htp]
\centering
    \centering
    \begin{tabular}{cc}
       (a) & (b)\\
        \includegraphics[width=0.235\textwidth]{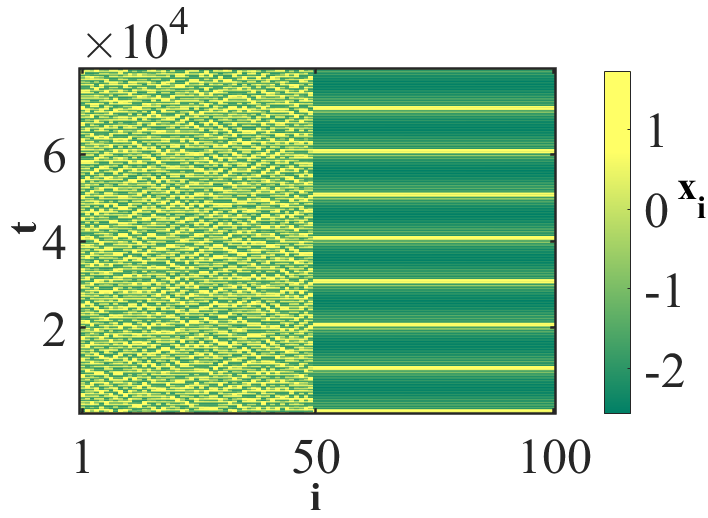} &
      \includegraphics[width=0.235\textwidth]{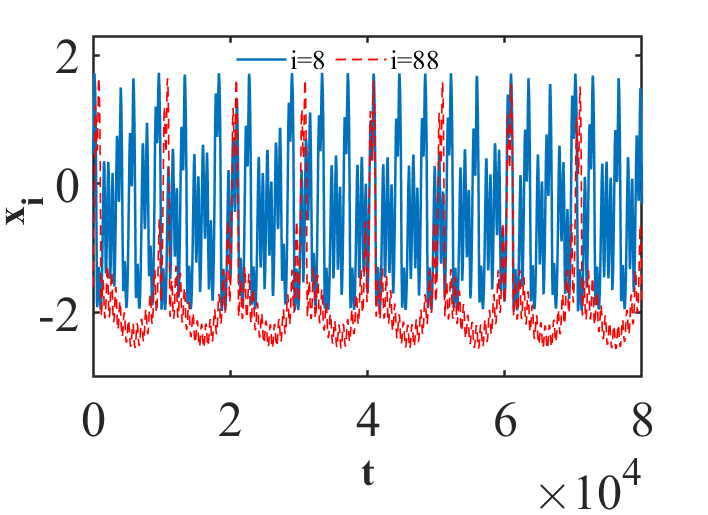}\\

    \end{tabular}
    \label{tab:placeholder}

\caption{Chimera-like state induced by a localized periodic field for $\epsilon=0$. (a) Spatiotemporal evolution of $x_i$ when $E_{\text{ext}}=1.5\sin(2\pi f t)$ is applied to neurons $i>50$ ($f=0.01$, $d=0.001$). (b) Time series of $x_i$ for neurons $i=8$ (no field) and $i=88$ (with field). The state is characterized by SI$=0.52$ and DM$=1$. Parameters: $k=0.001$, $\omega=1.004$, $r=0.007$, $N_E=50$.}
\label{fig:9}
 \end{figure}

With $\epsilon=0$, neurons interact only through electrical coupling. Figure~\ref{fig:9} shows the spatiotemporal pattern and time series when the field $E_{\text{ext}}(t)=1.5\sin(2\pi f t)$ is applied to neurons $i>50$ with $f=0.01$ and $d=0.001$. The stimulated neurons form a coherent domain with regular spiking, while the unstimulated neurons remain chaotic and incoherent, yielding a chimera-like state with SI$=0.52$ and DM$=1$.

\begin{figure}[htp]
\centering
    \centering
    \begin{tabular}{cc}
       (a) & (b)\\
        \includegraphics[width=0.235\textwidth]{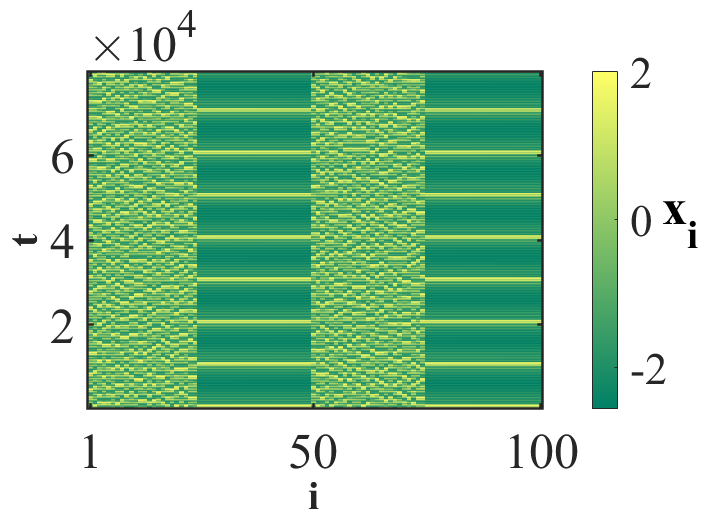} &
      \includegraphics[width=0.235\textwidth]{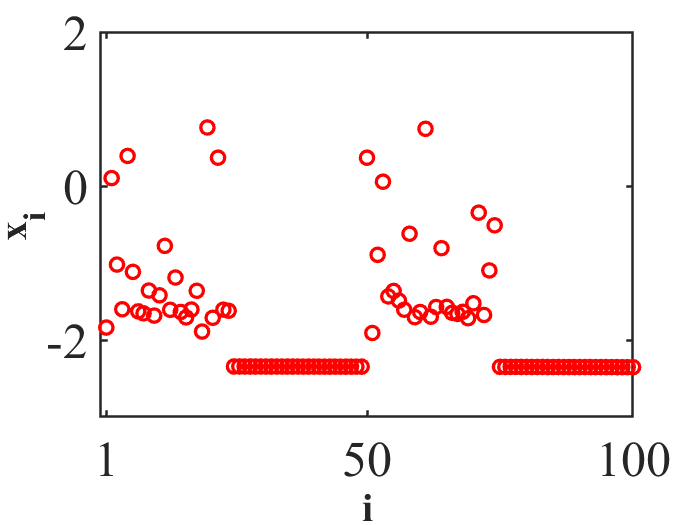}\\

    \end{tabular}
    \label{tab:placeholder}

\caption{Multichimera state induced by a two-region field for $\epsilon=0$. (a) Spatiotemporal dynamics and (b) snapshot of $x_i$ when the field is applied to neurons $25<i<50$ and $75<i<100$. Two coherent domains alternate with two incoherent domains, with SI$=0.65$ and DM$=2$.  Snapshot taken at $t=5000$(post-transient regime)}
\label{fig:11}
 \end{figure}

Applying the field to two disjoint regions generates multichimera patterns. In Fig.~\ref{fig:11}, neurons with $25<i<50$ and $75<i<100$ are stimulated; the spatiotemporal pattern and snapshot show two coherent domains alternating with two incoherent domains, with SI$=0.65$ and DM$=2$.

\begin{figure}[htp]
\centering
    \centering
    \begin{tabular}{cc}
        (a) & (b)\\
        \includegraphics[width=0.235\textwidth]{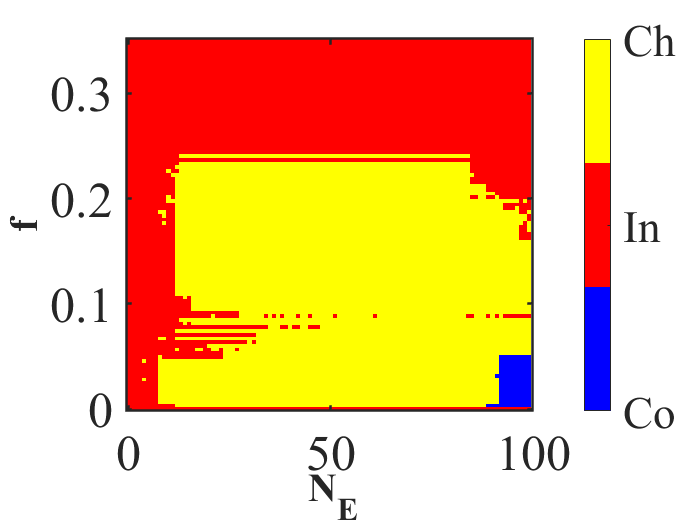} &  
       \includegraphics[width=0.235\textwidth]{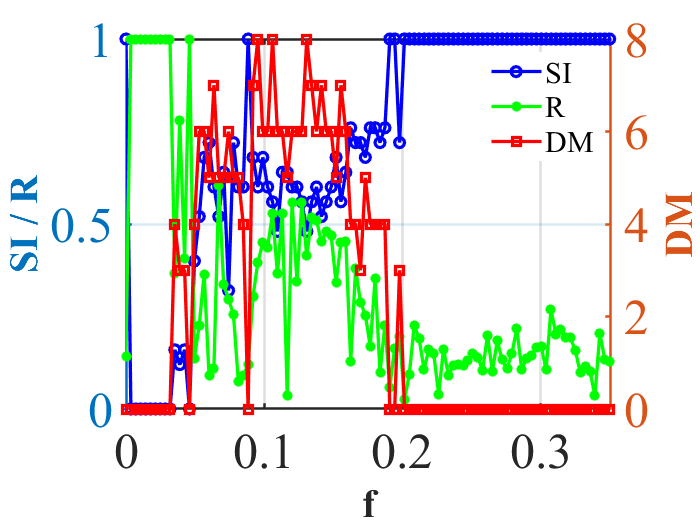}\\

    \end{tabular}
    \label{tab:placeholder}
    \caption{Field-induced regimes for $\epsilon=0$. (a) Color-coded phase diagram in the $(f, N_E)$ plane highlighting coherent (Co), incoherent (In), and chimera (Ch) regimes, identified using SI, $R$, and DM. (b) $SI$, $R$, and $DM$ as functions of $f$ for $N_E = 95$, revealing coherent states at low frequencies, chimera/multichimera states in the intermediate range ($0.05 <f <0.2$), and incoherent dynamics at higher frequencies.. Threshold $\delta=0.08$.}
\label{fig:12}
 \end{figure}

To summarize the dependence on field parameters, Fig.~\ref{fig:12}a shows coherent (blue, Co), chimera (yellow, Ch), and incoherent (red, In) states, classified using the Kuramoto order parameter, discontinuity measure, and strength of incoherence. Chimera states occupy the central region, incoherence dominates at high $f>0.25$ and low $N_E<10$, and full coherence is restricted to high $N_E>90$ and low $f<0.025$. Fig.~\ref{fig:12}b shows synchronization measures versus $f$ for $N_E=95$: Kuramoto order parameter $R$ (green), strength of incoherence $SI$ (blue, left axis), and discontinuity measure $DM$ (red, right axis). The coherent state ($f<0.05$) shows $R\approx 1$, $SI\approx 0$, $DM\approx 0$. Chimera states ($0.05<f<0.2$) are marked by oscillatory $DM$ with high peaks and fluctuating $R$. For $f>0.2$, the system becomes incoherent with $SI\to 1$ and $DM= 0$.

\begin{figure*}[htp]
\centering
   
    \begin{tabular}{ccc}
        (a) & (b) & c\\
        \includegraphics[width=0.235\textwidth]{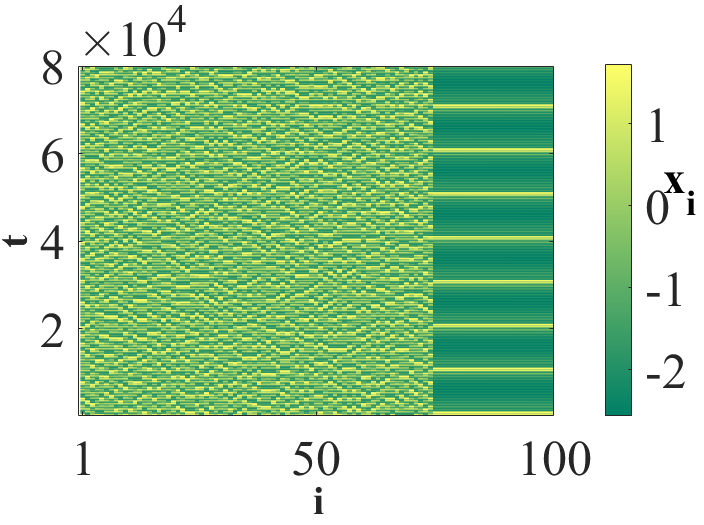} &  
       \includegraphics[width=0.235\textwidth]{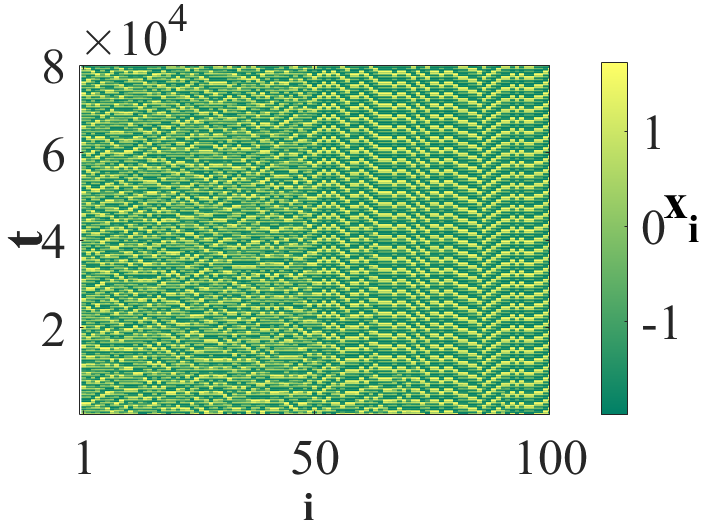} &
      \includegraphics[width=0.235\textwidth]{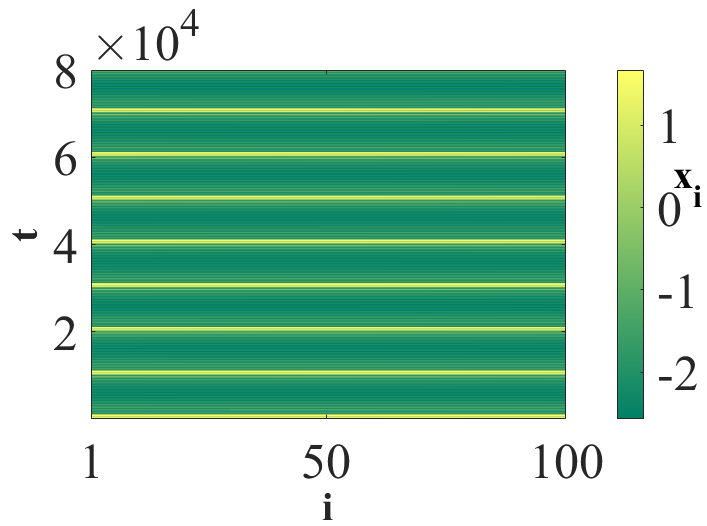}\\
       
    \end{tabular}
    \label{tab:placeholder}
\caption{Examples from the phase diagram in Fig.~\ref{fig:12}. (a) Single chimera state (SI$=0.8$, DM$=1$) for $N_E=25$, $f=0.01$. (b) Global incoherence (SI$=1$, DM$=0$) for $N_E=50$, $f=0.32$. (c) Complete synchronization (SI$=0$, DM$=0$) for $N_E=100$, $f=0.02$. Parameters: $d=0.001$.}
\label{fig:10}
\end{figure*}

Representative patterns from Fig.~\ref{fig:12} are displayed in Fig.~\ref{fig:10}. A single chimera (SI$=0.8$, DM$=1$) appears for $N_E=25$ and $f=0.01$ [Fig.~\ref{fig:10}(a)], global incoherence (SI$=1$, DM$=0$) arises for $N_E=50$ and $f=0.32$ [Fig.~\ref{fig:10}(b)], and full synchronization (SI$=0$, DM$=0$) is obtained when all neurons ($N_E=100$) are stimulated at $f=0.02$ [Fig.~\ref{fig:10}(c)].

\subsubsection{With chemical coupling}

\begin{figure}[htp]
\centering
    \centering
    \begin{tabular}{cc}
        (a) & (b)\\
        \includegraphics[width=0.235\textwidth]{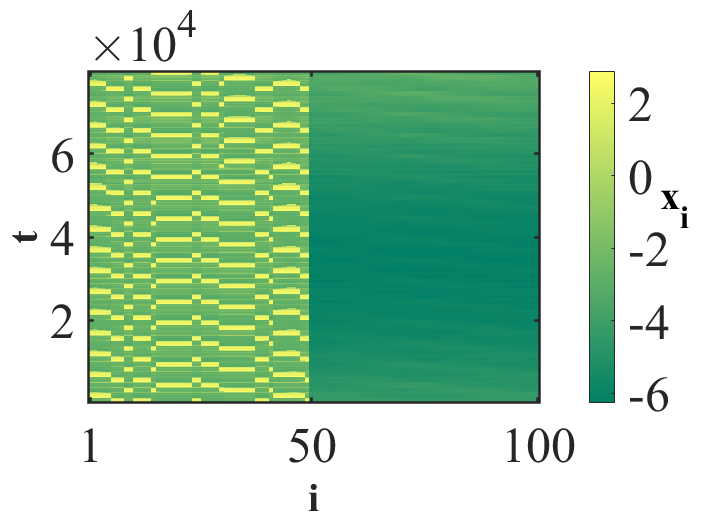} &  
       \includegraphics[width=0.235\textwidth]{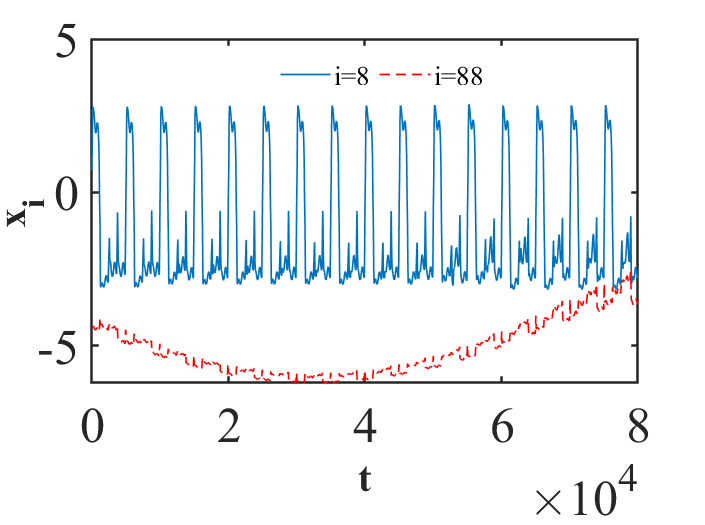}\\
             
    \end{tabular}
    \label{tab:placeholder}
\caption{Chimera state with hybrid coupling and localized field. (a) Spatiotemporal pattern for $d=0.001$, $\epsilon=10$, $r=0.007$, and $f=0.001$, with a coherent low-activity domain in the stimulated region. (b) Time series of neurons $i=8$ (no field) and $i=88$ (field), showing periodic spiking with small amplitude fluctuations versus suppressed dynamics. SI$=0.48$, DM$=3$. Parameters: $a=0.7$, $c=0.1$, $\xi=0.175$, $A=0.9$, $T=5$, $b=0.4$, $I=0.5$, $k=0.001$, $\omega=1.004$, $p=40$.}
\label{fig:13}
\end{figure}

We finally consider the combined effect of electrical and chemical coupling under external fields. Figure~\ref{fig:13} shows a chimera state for $d=0.001$, $\epsilon=10$, $r=0.007$, and $f=0.001$ when the field acts on one region. The stimulated neurons form a coherent low-activity domain, while the unstimulated neurons exhibit quasi-periodic dynamics, yielding SI$=0.48$ and DM$=3$.

\begin{figure*}[htp]
\centering
   
    \begin{tabular}{ccc}
        (a) & (b) & (c)\\
        \includegraphics[width=0.235\textwidth]{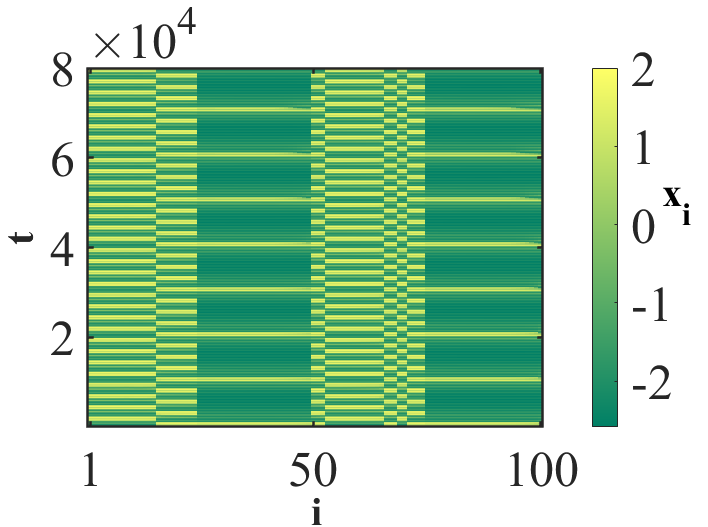} &  
       \includegraphics[width=0.235\textwidth]{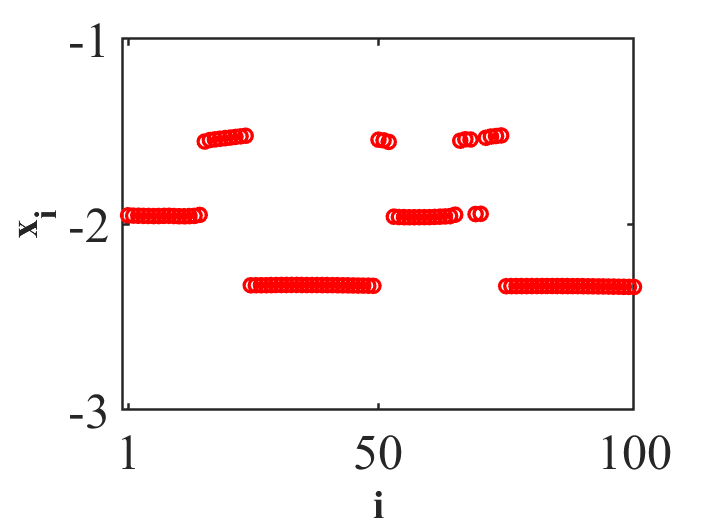}&      
       \includegraphics[width=0.235\textwidth]{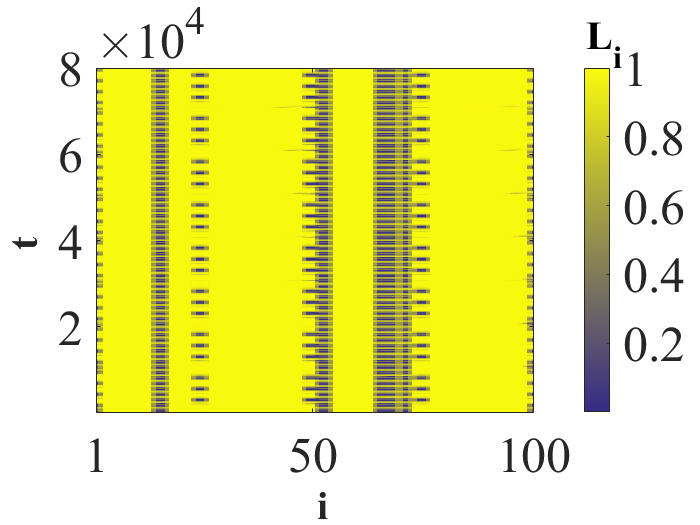}\\
       (c) & (d) & (d)\\
       \includegraphics[width=0.235\textwidth]{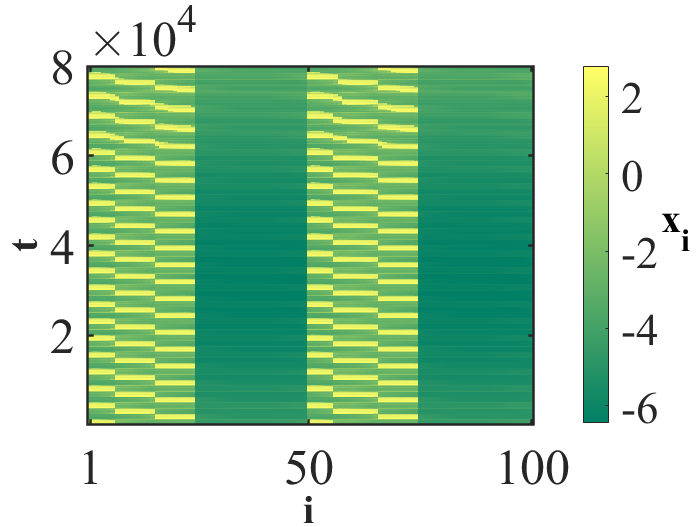} &
       \includegraphics[width=0.235\textwidth]{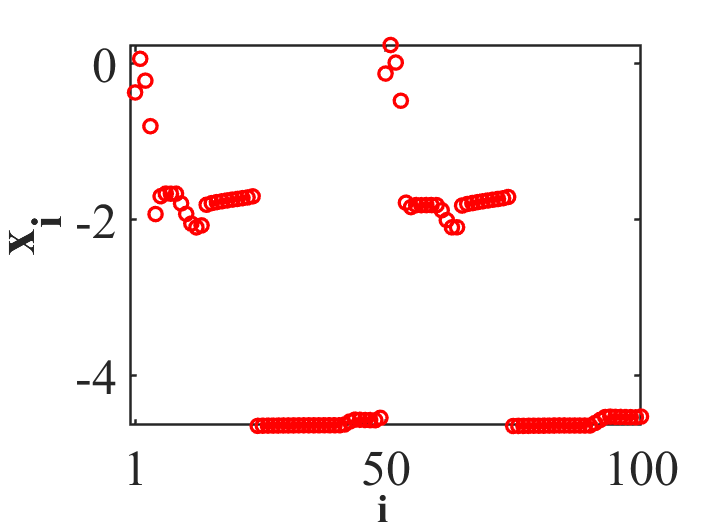}&
       \includegraphics[width=0.235\textwidth]{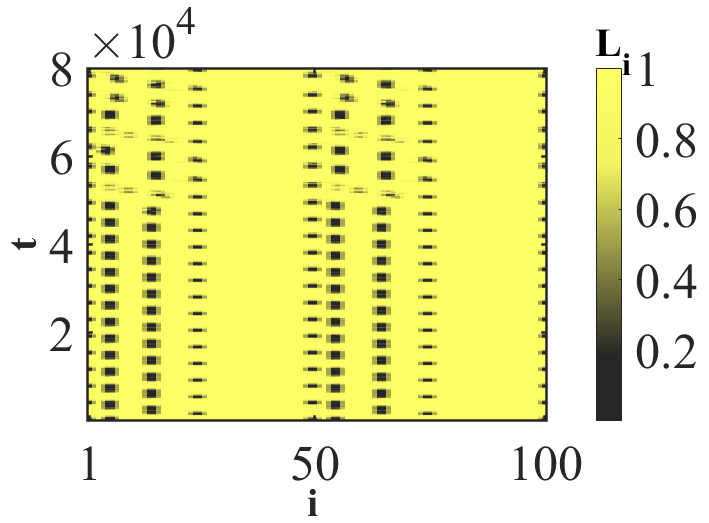}\\
        
    \end{tabular}
    \label{tab:placeholder}
\caption{Multichimera states in a hybrid-coupled network with multi-site external fields. (a–c) For $\epsilon=0.2$ and $f=0.01$, incoherent regions consist of multiple coherent clusters. (d–f) For $\epsilon=10$ and $f=0.001$, a different multichimera pattern emerges. The local order parameter $L_i$ in (c) and (f) highlights the alternation between coherent and incoherent domains. Parameters: $a=0.7$, $c=0.1$, $\xi=0.175$, $A=0.9$, $T=5$, $b=0.4$, $I=0.5$, $k=0.001$, $\omega=1.004$, $p=40$, $r=0.007$.}
\label{fig:14}
\end{figure*}

When the field is applied to two regions in the presence of chemical coupling, multichimera states with richer structure appear. Figure~\ref{fig:14} shows examples for weaker ($\epsilon=0.2$, $f=0.01$) and stronger ($\epsilon=10$, $f=0.001$) chemical coupling; in both cases, the spatiotemporal patterns and $L_i$ reveal multiple alternating coherent and incoherent domains, with incoherent regions composed of coherent clusters.

\section{Discussion}

This work analyzed a thermosensitive FitzHugh--Nagumo network with hybrid synaptic coupling and intrinsic electric fields under spatially localized, weak external electric fields. Numerical simulations revealed robust transitions between incoherent activity, coherent traveling waves, and chimera or multichimera states, controlled jointly by the chemical coupling strength, the cell-size–dependent intrinsic field parameter $r$, and the frequency and spatial extent of the applied field. Chemical nonlocal coupling was shown to be essential for the stabilization of traveling waves and traveling chimera patterns, whereas purely electrical nearest-neighbor coupling of chaotic thermosensitive neurons produced only incoherent network states.

Within this chemically organized backdrop, the external field frequency emerges as a critical control parameter. The pronounced impact of low-frequency fields ($f \lesssim 0.2$) compared to the weak influence of higher frequencies (Fig.~\ref{fig:12}) reflects an intrinsic timescale interaction in the model: membrane potential variables evolve quickly, whereas the ion-current variables $y_i$ and intrinsic field variables $E_i$ evolve on slower timescales. Low-frequency forcing can interact constructively with these slow variables, accumulating phase shifts and modulating their trajectories over many cycles, while high-frequency components average out and produce little net effect. This frequency-dependent modulation is consistent with earlier work on field-controlled chimera states in non-thermosensitive networks \cite{simo2021chimera}. This work's main contribution is to demonstrate how thermosensitive parameters and intrinsic field feedback alter the effective frequency range for control, going beyond the basic observation of frequency-dependent effects.

The collective dynamics of this thermosensitive network arise from the interplay of intrinsic properties like chaotic excitability and cell-size-dependent electric feedback, and external influences from hybrid coupling and applied fields. We observe three fundamental control mechanisms: (i) Wave generation via chemical coordination: in the absence of an external field, strong nonlocal chemical coupling is essential to override chaotic, electrically-driven incoherence and establish coherent traveling waves (Figs.~\ref{fig:4} and~\ref{fig:8}). Purely electrical coupling of chaotic neurons leads only to disordered activity (Figs.~\ref{fig:6}). (ii) Field-induced chimera without chemical coupling: when only electrical coupling is present, a localized, low-frequency external field can directly suppress chaos in the stimulated neurons, converting them to periodic firing. This contrast between the periodically coherent (stimulated) domain and the chaotic (unstimulated) background creates a chimera-like state (Figs.~\ref{fig:9} and~\ref{fig:10}). (iii) Field-modulated patterns with hybrid coupling: when chemical coupling is active, the same localized field can silence or reduce activity in the targeted region. This interaction carves out stable coherent domains of low activity within the network, resulting in structured multi-chimera states (Figs.~\ref{fig:13} and~\ref{fig:14}). The field's effectiveness is frequency-dependent: low frequencies reshape dynamics, while high frequencies leave the network unchanged (Figs.~\ref{fig:11} and~\ref{fig:12}).

The suppression of spiking in the stimulated region exemplifies this interaction. In the hybrid coupling regime, a localized low-frequency field strongly reduces activity in targeted neurons while leaving unstimulated regions chaotic or quasi-periodic (Fig.~\ref{fig:13}). Mechanistically, the low-frequency term $E_{\text{ext}} = E_{m}\sin(2\pi ft)$ enters the intrinsic field equation $dE_{i}/dt = ky_{i} + E_{\text{ext}}$, introducing a bias that shifts stimulated neurons toward low-activity trajectories. This phenomenological suppression illustrates how weak, localized fields can stabilize silent states in thermosensitive hybrid networks, though a detailed biophysical interpretation would require more realistic membrane dynamics.

From a broader perspective, the model contributes in three ways: it extends frequency-dependent control of chimera states to thermosensitive networks; it shows that intrinsic fields (via cell size $r$) modulate network susceptibility to external forcing; and it demonstrates that spatially targeted low-frequency fields can induce and control structured multichimera patterns, offering a conceptual basis for selective neuromodulation. Thermosensitivity, encoded in parameter $b$, further shapes this landscape by shifting boundaries between periodic and chaotic firing at the single-neuron level. Finally, the hybrid coupling architecture is essential: chemical synapses enable global pattern formation, while electrical coupling and intrinsic feedback regulate local excitability. Their interplay creates a hierarchical structure reminiscent of biological circuits, providing a minimal setting to study how different coupling channels cooperate under external stimulation.
These findings connect naturally with recent literature on thermosensitive neurons and chimera states. Hussain et al. \cite{hussain2021chimera} analyzed chimera formation in thermosensitive FitzHugh-Nagumo networks without external fields, demonstrating that the temperature coefficient $b$ strongly shapes the region in which chimera states occur, thereby establishing thermosensitivity as a key determinant of the intrinsic synchronization landscape. The present study extends this picture by introducing externally applied fields as additional control inputs, confirming that thermosensitive parameters continue to delimit synchronization regions even when the network is driven and showing how field parameters interact with this landscape. Ji and Mao \cite{ji2025chimera} examined chimera states in a ring network of thermosensitive FitzHugh-Nagumo neurons with nearest-neighbor memristive coupling, synaptic crosstalk, and explicit time delays, finding that variations in crosstalk intensity and delay can generate chimera, multichimera, chimera-death, and fully synchronized patterns. In their framework, thermosensitivity regulates single-cell excitability, while crosstalk and delays reorganize coherent and incoherent domains, without including weak macroscopic electric fields. The two approaches are complementary: Ji and Mao clarify how internally generated coupling effects sculpt chimera patterns in thermosensitive networks, whereas the present model focuses on how externally applied fields interact with hybrid synapses and intrinsic electric feedback, with intrinsic parameters selecting which field frequencies and stimulated fractions of neurons yield incoherence, chimera-like states, or global synchrony.

This study employs a simplified yet informative model to explore how weak electric fields modulate spatiotemporal patterns in thermosensitive hybrid-coupled networks. Several simplifying assumptions were made to keep the system tractable, including: a ring topology; identical neurons; a phenomenological intrinsic field; a static, sigmoidal chemical synapse model; the omission of noise; and a fixed temperature. While these choices limit direct biological realism, they allowed us to isolate and characterize fundamental control mechanisms in a clean mathematical setting. Future work should extend this framework by incorporating more realistic features, such as heterogeneous neurons, small-world topologies, synaptic plasticity, dynamic temperature, biophysically detailed ephaptic coupling, and stochastic noise, to bridge the gap between these conceptual findings and applications in neuromodulation and bio-inspired computing.

\section{Conclusions}
This work demonstrates that weak, spatially targeted electric fields can effectively control collective dynamics in thermosensitive neuronal networks with hybrid coupling. We have shown that strong nonlocal chemical coupling is necessary to organize chaotic, electrically coupled neurons into coherent traveling waves. In the absence of chemical coupling, a localized low-frequency field can directly suppress chaos in stimulated neurons, inducing chimera-like states by creating a contrast between periodic (stimulated) and chaotic (unstimulated) domains. When combined with chemical coupling, the same field can silence activity in targeted regions, leading to structured multichimera patterns. The effectiveness of the field is frequency-dependent: low frequencies reshape network dynamics, while high frequencies have negligible impact.

Importantly, the intrinsic electric field (modulated by cell size $r$) and thermosensitive parameters (via $b$) reshape the network's susceptibility to external control, defining the effective frequency window and parameter ranges for pattern selection. These findings provide a principled framework for understanding how endogenous properties and external stimulation interact to generate complex spatiotemporal activity, with implications for the design of selective neuromodulation strategies and bio-inspired computing systems.

\section*{Acknowledgments}

ACR thanks the São Paulo Research Foundation (FAPESP). FFF and ACR thank the Brazilian National Council for Scientific and Technological Development (CNPq). ELFN thanks the Brazilian Federal Agency for Support and Evaluation of Graduate Education (CAPES).

\section*{Funding declaration}

This work was supported by the São Paulo Research Foundation (FAPESP: grant 2013/07699-0 and grant 2025/18142-3), the Brazilian National Council for Scientific and Technological Development (CNPq, grants 303359/2022-6 and 316664/2021-9), and the Brazilian Federal Agency for Support and Evaluation of Graduate Education (CAPES, grant 001).

\section*{Conflict of interest}

The authors declare that they have no conflict of interest.

\bibliographystyle{unsrt}
\bibliography{references}

\end{document}